\documentclass[12pt]{iopart}
\usepackage[latin1]{inputenc}
\usepackage{graphicx}
\usepackage{bm}
\usepackage{subfigure}
\usepackage{latexsym}
\usepackage{graphics}
\usepackage{dcolumn}
\usepackage{bm}
\usepackage{amsfonts}
\usepackage{amssymb}
\DeclareGraphicsExtensions{.pdf,.png,.jpg}
\usepackage[comma,numbers,sort&compress]{natbib}

\def\Agsqrt{Ag-$\sqrt{3}~$}
\def\77 {$(7 \times 7)~$}

\newcommand{\gradC}{\textdegree C }

\begin{document}
\title{Tuning of one-dimensional plasmons by Ag-Doping in Ag-$\sqrt{3}$-ordered atomic wires}
\author{U Krieg, Yu Zhang, C  Tegenkamp, H Pfn\"ur}
\address{Institut f\"ur Festk\"orperphysik, Leibniz--Universit\"at
Hannover, Appelstrasse 2, 30167 Hannover, Germany}
\ead{pfnuer@fkp.uni-hannover.de}

\begin{abstract}
We generated arrays of silver wires with a height of one atom and an average width of 11 atoms 
on the Si(557) surface via self assembly with local $\sqrt{3}\times\sqrt{3}$ order, 
and investigated the 1D plasmon formation in them using a combination 
of high resolution electron loss spectroscopy with low energy electron diffraction.  
As it turned out by a series of thermal desorption experiments followed by adding small concentrations
of Ag, pure \Agsqrt ordered arrays of nanowires, separated by (113) facets, 
are intrinsically semi-metallic or semiconducting, i.e. metallicity of the Ag wires seems to be caused by 
excess atoms added to the (locally) perfectly ordered $\sqrt{3}\times\sqrt{3}$ layer.  
The proof has been carried out by postadsorption of Ag atoms in the range between 
0.004 to 0.03 monolayers and the quantitative determination of the frequency dependence 
of the 1D plasmon due to this excess Ag concentration.  
As expected for a doping mechanism, there is no minimum excess concentration. The lack of temperature 
dependence is not compatible with formation of an adatom gas in the second layer, but suggests extrinsic
doping by adatoms bound at the stepped (113) facets. 
Although strong deviations from a nearly-free electron gas are expected in 1D, 
the Ag concentration dependence of the 1D plasmonic losses is 
fully compatible with the $\sqrt{n_e}$ dependence predicted by this model. 
Adsorption of traces of residual gas can have a qualitatively similar doping effect.  
\end{abstract}
\submitto{\NJP}
\maketitle

\section{Introduction}
Coupling of energy into collective excitations like plasmons in metallic nanostructures, particularly
of light, has been in the focus of interest over many years. ``Conventional'' surface plasmons of 3D bulk 
metals form plasmon polaritons with an incident electromagnetic field and allow formation 
of wave guides as well as localization of the excitation in sub-wavelength structures \cite{Barnes03,Maier07}. 
A fascinating application is the formation of surface plasmon coupled lasers in microcavities \cite{Bergman03,Marell2011}. 

The plasmons of ultra-thin metallic sheets or of wires with a few atoms in diameter, forming 2D sheet 
plasmons or 1D wire plasmons, have additional attractive properties. 
Due to their flat dispersion, which starts at zero energy for large wavelengths, much shorter wavelengths
(below 10 nm) can be achieved compared with surface plasmons, allowing for better localization, 
as demonstrated, e.g. in graphene nanostructures \cite{Yan13}, whose 2D plasmonic properties were 
also studied in detail \cite{Tege11,Pfnuer11}. Shielding of the plasmonic excitation 
for 2D sheet plasmons by a metal substrate or interaction of stacks results in linearization of the 
dispersion \cite{Silkin04,Morawitz93}, which for an unshielded single metallic 
sheet would start as $E \sim\sqrt{k_{\|}}$. $k_{\|}$ is the in-plane wavevector. 
The linearization of dispersion allows for distortionless 
signal transport, and is particularly attractive, therefore. 
Quasi-linear dispersion of these acoustic surface plasmons 
has been demonstrated for several surfaces of metals such as Au, Cu, and 
Be \cite{Diaconescu07,Park10,Pohl10,Jahn12,Vattuone13}. 
All of them have Shockley surface states that cross the Fermi level.  

\begin{figure}
\begin{center}
{\large\bf a)} \includegraphics[width=0.5\textwidth]{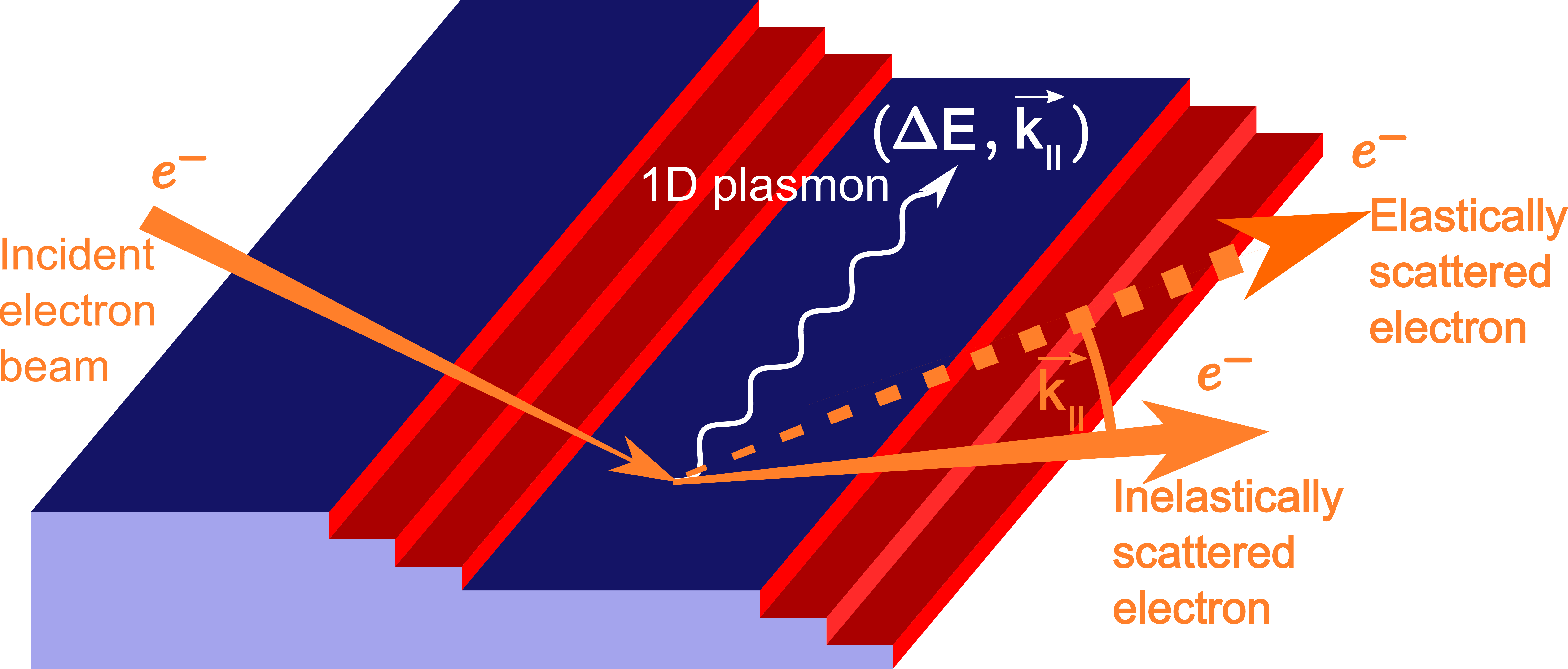}\\
{\large\bf b)} \includegraphics[width=0.7\textwidth]{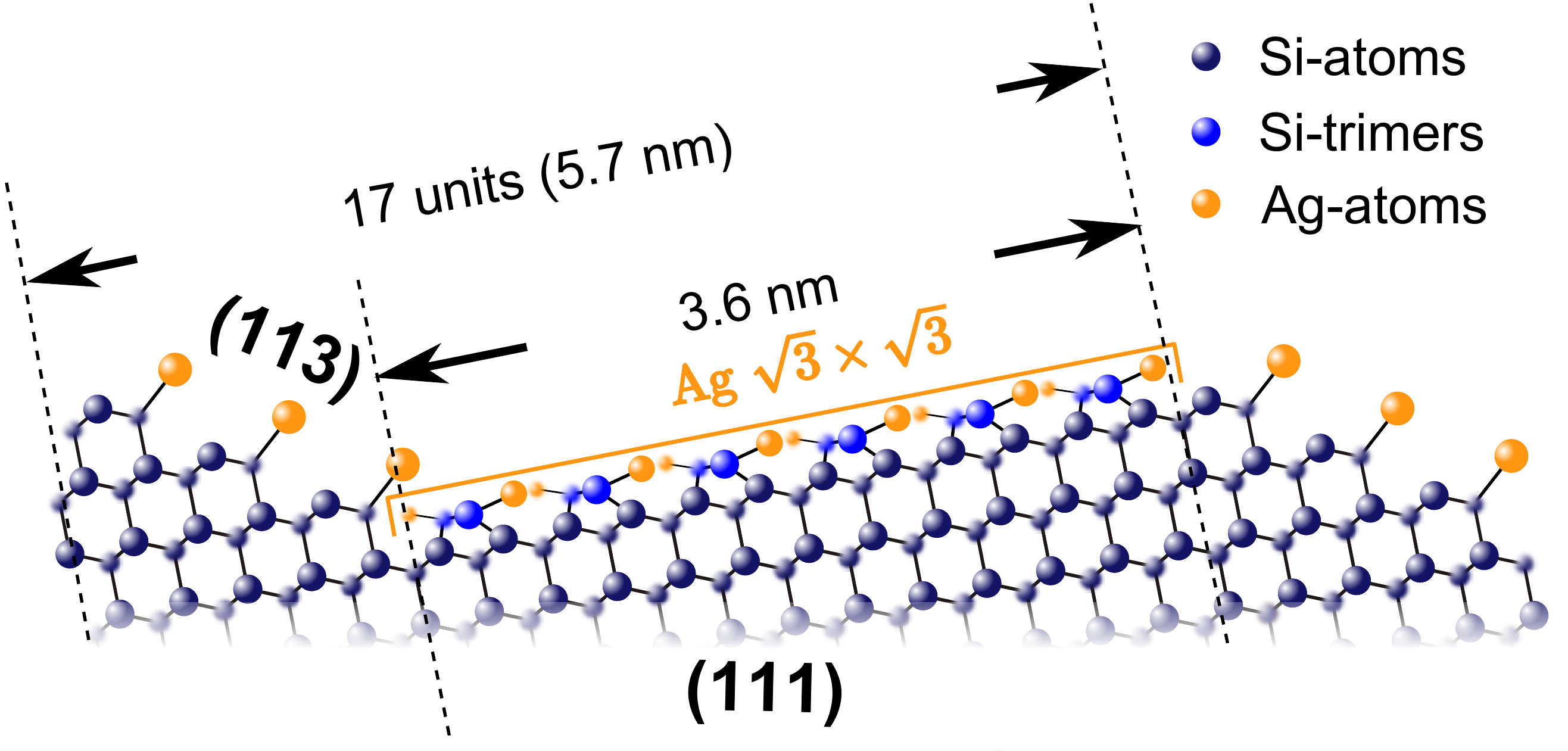}
\end{center}
\caption{\label{concept} a) Schematic of inelastic scattering of electrons at an array of atomic conducting wires (blue) separated by three steps, which form a mini-facet as in Si(557) and electrically separate the wires. Since also the substrate is assumed to be insulating, plasmons can only travel along the wires.
b) Side view of the atomistic structure on Si(557), covered with a monolayer of Ag. Compared with the clean 
Si(557) surface, the mini-facets are reoriented to (113) orientation, and the (111) terraces are widened (see ref.~\cite{Krieg13}). The location of the Ag atoms at the steps is only drawn schematically.  
}
\end{figure}

Switching to 1D wire plasmons with their already built-in directionality of energy transport, 
the dispersion, according to theory, is intrinsically quasi-linear, since it starts at long wavelengths 
as  $E \sim k_{\|}\ln{k_{\|}}$ \cite{DasSarma85}. Surprisingly little is known about these plasmons, 
although this type of dispersion was corroborated by simulation \cite{Inaoka05} and by few experiments
\cite{Ruge10,Nagao07,Block11}. These partly also demonstrate dimensional cross-over to two dimensions \cite{Block11}. The general concept of coupling energy from electrons via inelastic scattering into this type of 
plasmons is sketched in Fig.~\ref{concept}a, which can act as plasmonic waveguides.

Recently, we presented results on 1D plasmon excitation in wires of $\sqrt{3}$-ordered Ag on Si(557)
\cite{Krieg13}. Although a coverage of one monolayer (1ML) was deposited, 1D dispersion was found, indicating electrical separation between quite 
densely packed atomic wires. The side view of the atomistic arrangement, using the honeycomb-chained trimer model for the (111)-terraces \cite{Wan93}, is shown fig.~\ref{concept}b). Please note that the position of the Ag atoms on the (113) facets is not known and only sketched schematically.
Interestingly, neither has the concentration dependence of the plasmon energy been tested 
for such 1D systems nor is the origin of conductance clear. 
For tests answering these questions the \Agsqrt phase on the stepped Si(557) surface turned out to be 
well suited. Since no angular resolved photoemission (ARPES) photoemission data for this system are 
available yet, the analogy with the similar Ag/Si(111) system tells that 
the perfect \Agsqrt phase might also be a semi-metal, as found for the monolayer of Ag on Si(111)
\cite{Ding91,Zhang01,Crain05}, and tuning of the plasmon resonance by Ag concentration may be possible \cite{Liu09}. Here the Ag-modified surface state, $S_1$, touches the Fermi level
from above, but remains empty. Only by a surplus of Ag atoms, charge is transferred 
to this state so that it gets partially filled, leading also to the 
observability of 2D sheet plasmons \cite{Nagao01}. 
In a recent publication \cite{Krieg13}, we showed that a similar phase is 
formed as $\sqrt{3}$-ordered wires of finite width on the (111)-oriented 
mini-terraces of the Si(557)-surface. Whereas a different kind of wires is 
formed at low coverage, $\Theta < 0.3$ ML, which turned out to be semiconducting, 
metallic conductance was found to be associated with the appearance of 
$\sqrt{3}$-ordered domains, as we demonstrated by high resolution electron energy loss spectroscopy 
(HREELS). A clear plasmonic loss was identified for this phase, which showed dispersion only in the 
direction along the wires. No dispersion, but only subband
formation and intersubband-plasmon excitation was observed in the direction normal to the wires. 

This metallicity is indeed caused by Ag atoms added to the perfect \Agsqrt phase, as 
we demonstrate in this paper. It seems that a certain surplus of Ag atoms is always 
generated during the standard preparation 
procedure resulting in a conducting array of wires. As we show below, this excess concentration 
of Ag atoms can be eliminated by an annealing treatment without removal of the $\sqrt{3}$-ordered 
wires. This means that also for this stepped surface a mechanism seems 
to be operative that is similar to the Ag/Si(111) system. On the other hand, this mechanism 
also allows to vary the concentration of charge carriers. Here we again use the plasmonic losses
as sensor, which allows to test the dependence of these losses on the Ag excess concentration. Since we 
are working in the low excess coverage regime ($\Delta\Theta < 0.05$ ML), it seems plausible that 
a linear relationship between excess Ag concentration and charge carrier concentration,$n_e$, holds. Thus we 
are able to test the dependence of the plasmon dispersion on $n$ in 1D via doping experiments.

\section{Experimental setup}
The plasmons were measured by using  a combination of high resolution electron loss spectrometer 
(EELS) as electron source with a low energy electron diffraction system (LEED) providing 
simultaneously high energy and momentum ($k_{||}$) resolution \cite{Claus92,Nagao00}. 
Typical operating parameters were 25 meV energy resolution at a $k_{||}$ resolution 
of $ 1.3 \times 10^{-2}~\AA^{-1}$. In order to check the surface structure after 
in-situ cleaning and after Ag adsorption profile analysis in LEED (SPA-LEED) was 
used. Ag was evaporated by electron-beam heating from a Molybdenum crucible. The 
Ag flux was controlled by a quartz microbalance located at the evaporator. This 
micobalance was calibrated with another one located at sample position, supplemented 
by calculated estimates and by STM. The different approaches agreed within 8\%, but the direct microbalance 
calibration was considered to be the most reliable. The uncertainty of the latter was less than 5\%. 
The coverage below is given with respect to the surface density of Si(111),  
i.e. 1~ML$\cong 7.84 \times 10^{14} cm^{-2}$.
The sample was prepared by repeated annealing cycles to 1100\gradC, which consisted 
of rapid heating to this 
temperature, a quench to 900\gradC, a linear cooldown to 800\gradC over 2 min before the sample was cooled 
down without further heating. Surface quality was subsequently checked with LEED.  
In order to determine accurately the peak positions of the plasmon
losses, the spectra have been fitted. The elastic peak was modeled by a Gaussian peaks and an exponentially decaying Drude tail while
the individual losses were modeled using an exponentially modified Gaussian (EMG) functions and a constant background. 

\section{Results and discussion}

\subsection{The reference \Agsqrt structure}
We first describe the generation of a \Agsqrt reference structure. Here, the aim was to 
generate large and perfectly ordered $\sqrt{3}$-terraces with as little excess 
Ag concentration as possible on them.  For our investigations, there is no need, however,  
that they cover the surface completely, since all parts of the surface not covered by this structure 
are semiconducting. Therefore, they do not contribute to the plasmon signal. 

\begin{figure}
 \centering
 \subfigure[1.2 ML of Ag, as prepared]
 {\includegraphics[width = 0.4\textwidth]{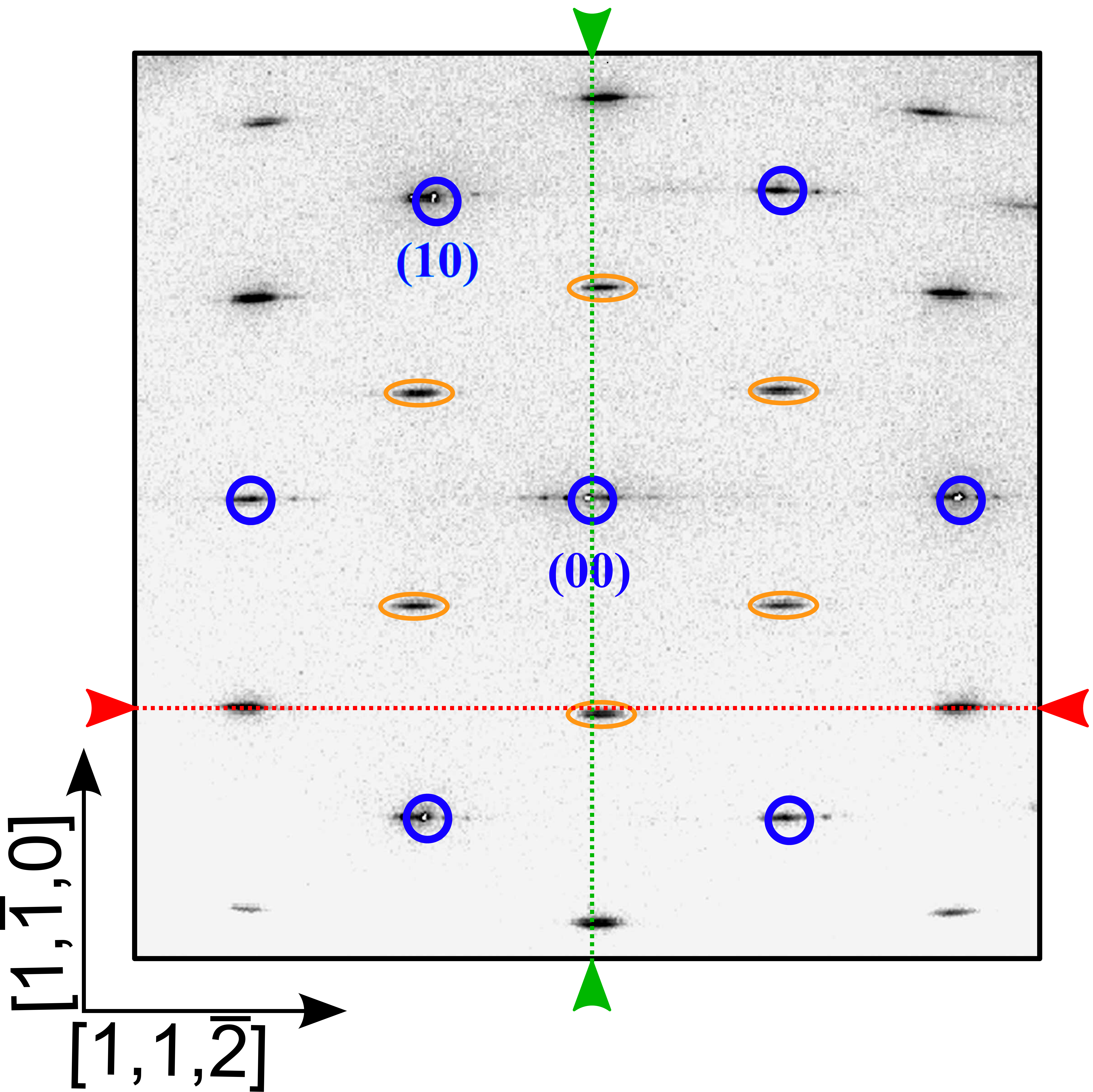}}\qquad
 \subfigure[after 2 heating cycles ]
 {\includegraphics[width = 0.4\textwidth]{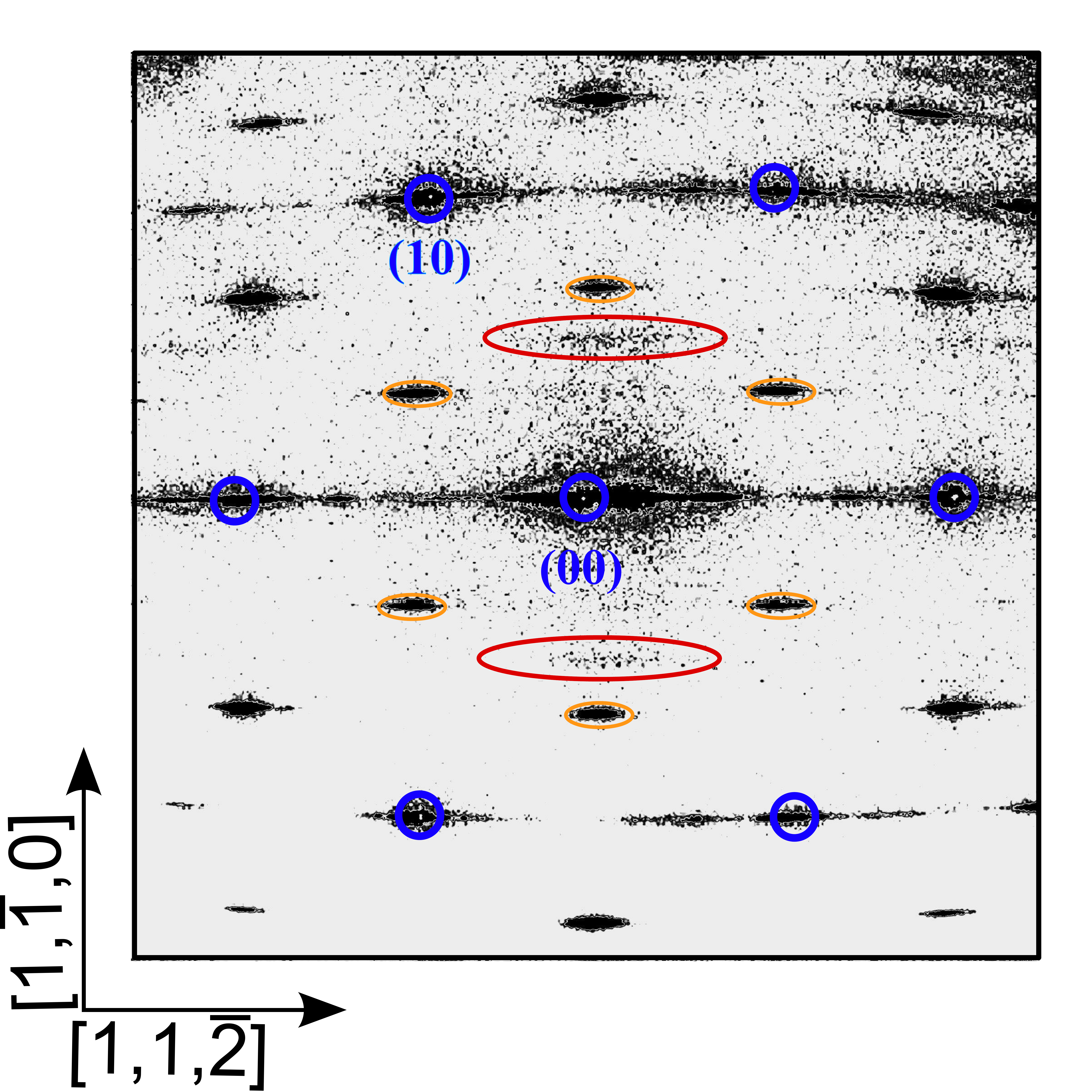}}\\
 \subfigure[line profiles]{ \centering 
 \includegraphics[width = 0.45\textwidth]{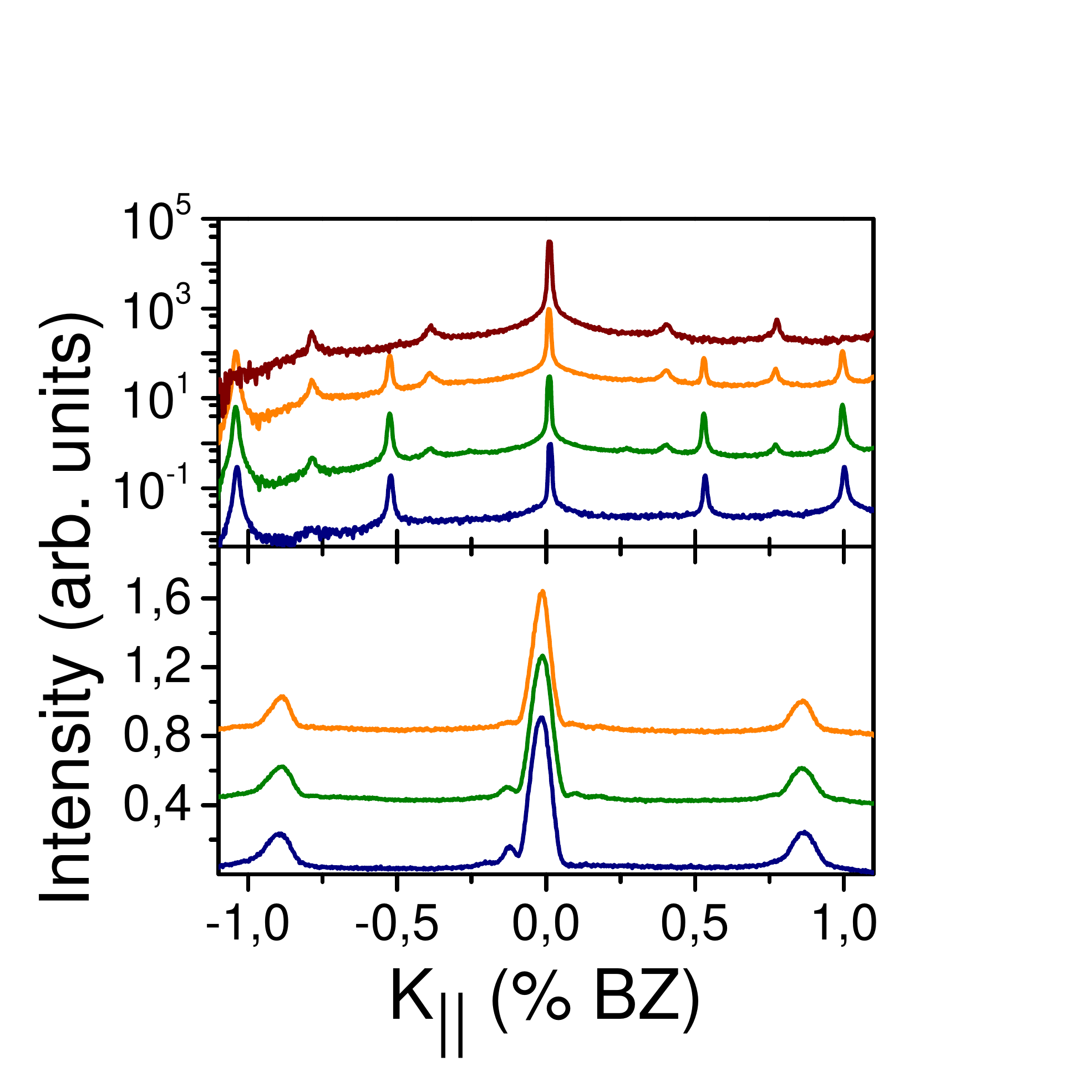}}\qquad
 \subfigure[Change of integrated LEED intensities]
 { \centering \includegraphics[width = 0.45\textwidth]{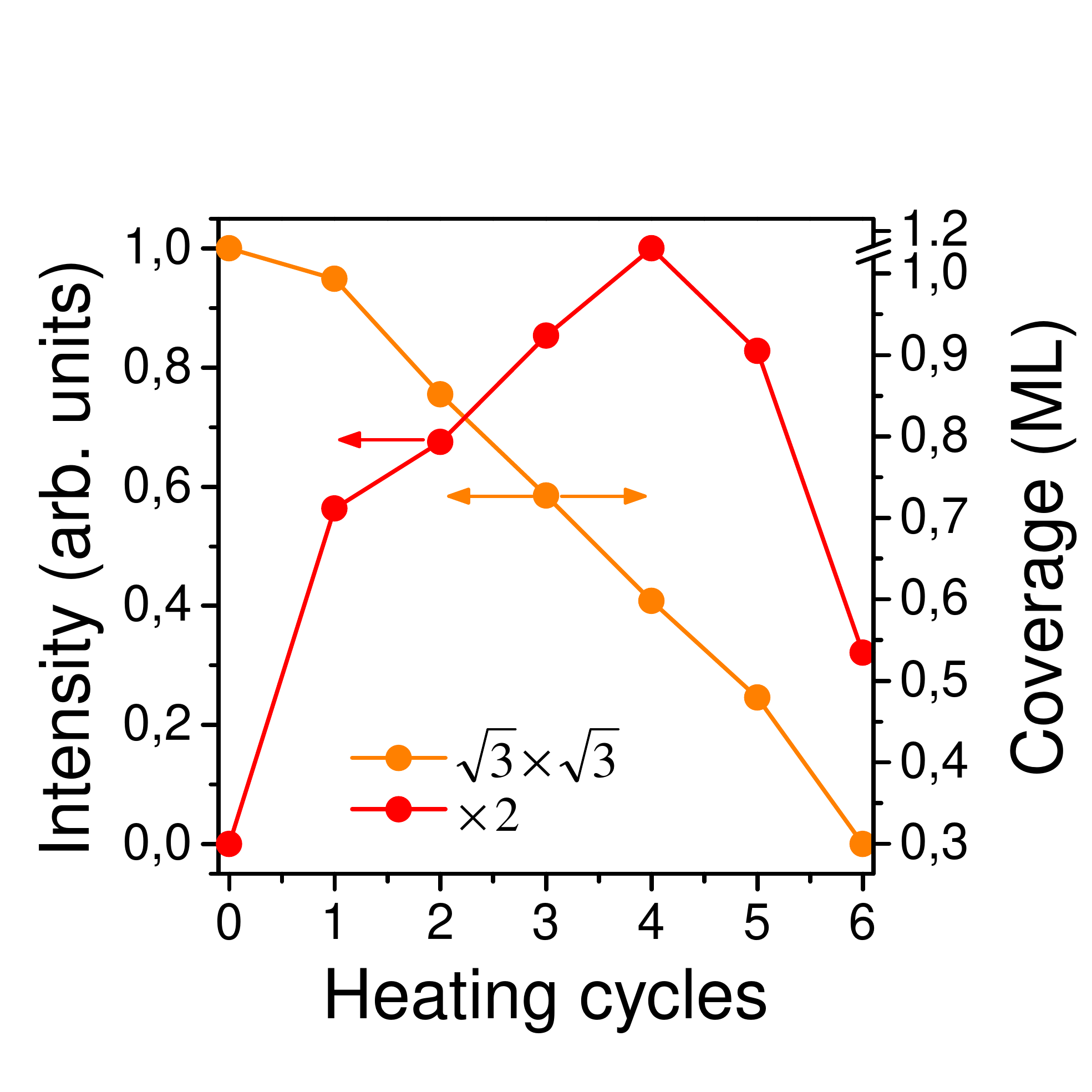}}
 \caption{SPA-LEED measurements. a) LEED pattern of a freshly prepared sample with 1.2 ML Ag. b) same after 
 2 heating cycles to 600\gradC. Additional $\times 2$ streaks appear due to local formation of silver-induced 
 atomic (semiconducting) wires \cite{Krieg13}. c) Line scans along the directions indicated by red and green 
 arrows in a), and after the zero (blue), two (green), four (orange) and six heating cycles.
 d) Change of integral LEED intensities of the \Agsqrt and $\times 2$ spots, depending on the number of 
 heating cycles. The data was normalized to the initial intensities of a freshly prepared sample. \label{fig:LEED_Control}}
 \end{figure}

In a first step we evaporated 1.2ML Ag onto the sample 
held at 500\gradC. This results in formation of 
a $\sqrt{3}\times \sqrt{3}$R30$^\circ$ structure on the (111)-oriented terraces, which is completed at one monolayer coverage, with the typical elongated spots characteristic of 
stripes with this order and an average width of 3.6 nm \cite{Krieg13} (see Fig.~\ref{fig:LEED_Control}a).
The 1.2ML Ag amount was chosen in order to start with a saturated $\sqrt{3}$ structure and to be able to then 
carry out controlled desorption experiments \cite{Wan93,Ueno03}. The excess coverage on the terraces and that 
on the (113) facets remains invisible in LEED, i.e. it is distributed randomly in very small islands, which 
were visible in STM. 

In a second and crucial step, the Ag covered surface was heated periodically to 600\gradC for 60s via direct current parallel to the steps 
with a 10min cooldown. This procedure leads to partial desorption of Ag, but at the same 
time to complete removal of metallicity, as shown below, without losing the majority of $\sqrt{3}$-ordered 
terraces.

After heating cycle we performed SPA-LEED measurements (Fig. \ref{fig:LEED_Control}) to check the surface. The first heating process was controlled with a Pyrometer during which the current-voltage curve was measured. 
Using this information we were able to perform the same procedure at the same spot after exactly the same heating steps. The partial 
desorption of Ag without destruction of the \Agsqrt structure is obvious from Fig.~\ref{fig:LEED_Control}b)
\cite{Ueno03,Crain05,Uhrberg02}. Whereas the $\sqrt{3}$ superstructure remains intact after the heating cycles,  its intensity is reduced, but the peak widths remain constant until the spots are close to disappearance. 
As can be estimated from the reduction of (integrated) LEED  intensities 
in the $\sqrt{3}$-superstructure spots (Fig.~\ref{fig:LEED_Control}d) that the  
loss of Ag coverage is  about 0.15 ML per cycle. 
Thus the \Agsqrt reconstruction is still covering the surface to about 85\% after two cycles. This means 
that the surface consists of finite domains of $\sqrt{3}$-structure and semiconducting thin wires.  
After 6 heating cycles  no \Agsqrt spots remain, but only the $\times 2$ streaks. 
There is no indication for reappearance  of the  $7\times 7$ reconstruction, i.e. the whole surface 
is still covered with the low-coverage species of silver induced atomic (and semiconducting) wires. 

\begin{figure}
 \begin{center}
  \includegraphics[width = 0.5\textwidth]{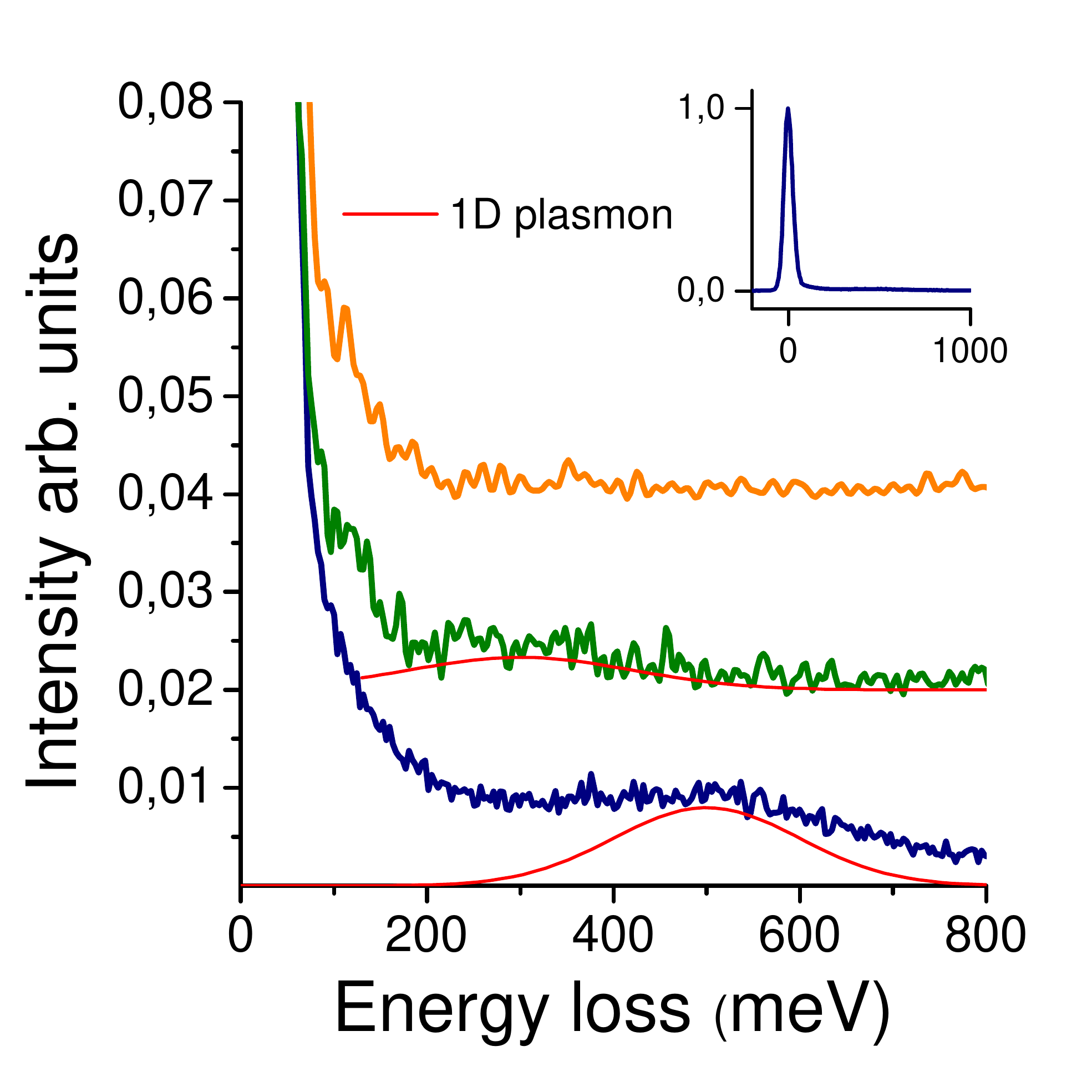}
 \end{center}
\caption{EEL-spectra after adsorption of 1.2 ML of Ag on Si(557) (bottom) and after sequential 
heating cycles to 600 \gradC for 60s (2nd and 3rd spectrum from bottom). They were measured at $k_\parallel = 6$\%BZ 
parallel to the steps at 20eV incident electron energy. The red 
lines denote schematically the plasmon losses at the positions and with the halfwidths 
obtained by fits. For better visibility, the spectra have been shifted. 
Inset: normalized elastic peak. \label{fig:EELS_Cleaning} }
\end{figure}

At the same time, the charge carrier concentration of Ag in the \Agsqrt stripes is reduced, as seen 
by the HREELS spectra shown in Fig.~\ref{fig:EELS_Cleaning}. Already after the second heating cycle the 
plasmon cannot be detected any more. In fact, upon further heating cycles the loss spectra remain 
unchanged, and their shape is identical to that measured for an Ag concentration of 
0.3 ML \cite{Krieg13}. The latter concentration marks the onset of $\sqrt{3}$-island formation as a function of coverage, as detected by LEED. Below this concentration only semiconducting nanowires were observed both on the 
(111) and (113) facets \cite{Krieg13}, as already mentioned.
These wires on both the (111) terrraces and on the steps are characterized by a double 
periodicity with respect to the Si substrate, which leads to $\times 2$ - streaks in the LEED pattern (see 
Fig.~\ref{fig:LEED_Control}b)) \cite{Krieg13,Morikawa08}. 
During the heating cycles described above these streaks 
reappear already after the first heating cycle, as seen from Fig.~\ref{fig:LEED_Control}c) and d). 
They are a signature 
of phase coexistence of $\sqrt{3}$- and atomic wire phases, which ends after six heating 
cycles. At this coverage only the semiconducting wire phase is left.
The strongly non-monotonic intensity change of the $\times 2$ streaks with 
a clear maximum after four heating 
cycles, however, cannot be explained by simple phase coexistence. It means that strong rearrangement mainly at the steps must occur during the first heating cycles, most likely due to preferential desorption of Ag at the steps.  

For the remaining description of our experiments 
we concentrate on the properties of Ag layers after two heating cycles, which turned out to be sufficient
to remove metallicity, as detectable by characteristic plasmon losses.
From the measurements shown in Fig.~\ref{fig:EELS_Cleaning} we can get an upper limit of the 
remaining doping of the \Agsqrt nanowires. The detection limit of a plasmon, using quantitative 
fits of the loss spectra, is about 100 meV. As deduced from the dispersion measured in 
ref.~\cite{Krieg13}, an electron concentration of $1.9\times 10^7$cm$^{-1}$ was found there. 
This resulted in a loss peak at 878 meV at 10\%BZ. 
Assuming $E_{loss}(k) \propto \sqrt{n_e}$ ($n_e$ is the electron density), we reduced the 
electron concentration by at least a factor of  $\left(878/100\right)^2 \approx 80$, i.e. below  
$2.4\cdot10^5$cm$^{-1}$. 

Summarizing this section, we have found a way to bring back the \Agsqrt layer into a semi-metallic or even 
semiconducting state, i.e. we have removed the Ag species below detection limit that is responsible for 
charge transfer into the Ag induced surface state. This was done at the expense of formation of an incomplete 
monolayer of \Agsqrt. We note here that the re-appearance of mobile charge 
carriers, as detected by EELS-LEED recording the plasmonic losses, happens already at 
the smallest amounts of additional Ag, i.e. without any threshold in added Ag concentration. 
This will become evident from our experiments with Ag addition described below. Evenly 
important, our findings do not depend on the exact amount of remaining \Agsqrt concentration after the 
heating cycles, i.e. they are a unique property of the \Agsqrt structure. 

\subsection{Plasmon loss induced by residual gas}\label{sec:time}

In the following we demonstrate that after the treatment of the \Agsqrt layer, prepared as described above,
with the complete removal of the characteristic plasmonic losses we are able to reverse this 
process both by adsorption of small amounts of Ag and even by adsorption of residual gas, i.e. 
we carried out ``doping'' of the \Agsqrt layer. 
We first describe the effects by 
residual gas adsorption, since the influence of residual gas has to be taken into account for the 
quantitative determination of loss frequencies as a function of Ag concentration.

\begin{figure}
 \centering
 \subfigure[]{\includegraphics[width = 0.4\textwidth]{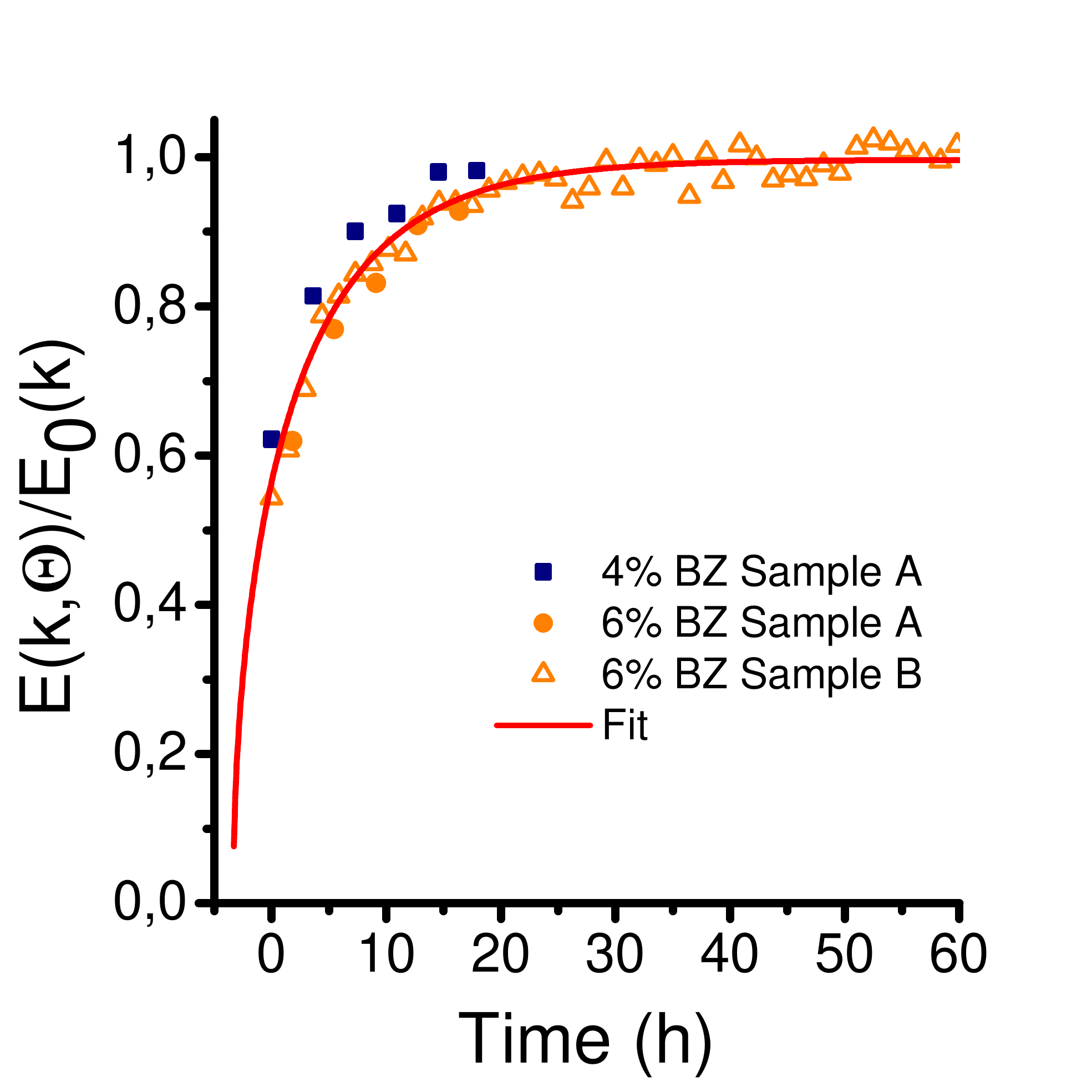}}\qquad
 \subfigure[]{\includegraphics[width = 0.4\textwidth]{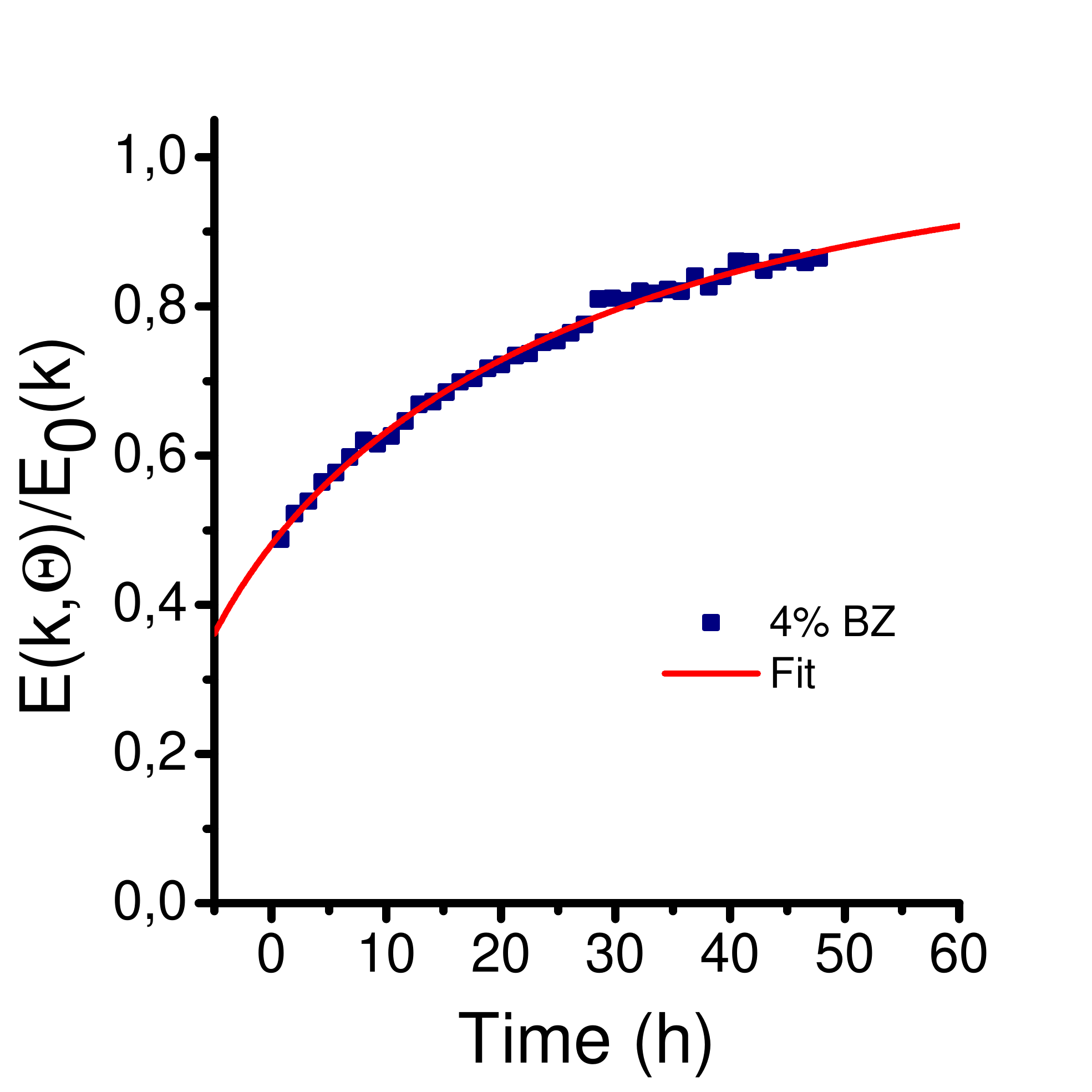}}\\

\caption{Change of the plasmon loss as a function of time due to adsorption of 
residual gas by the background pressure of $1\times 10^{-10}$ mbar at room temperature and at constant 
$k_\parallel$ after preparation of a 1.2 ML Ag layer on Si(557) and after two heating cycles. 
All data were normalized by the saturation value at a specific $k_\parallel$. Different colors of the 
data points correspond to different $k_\parallel$ values. Red lines: Fit of the data with the model 
of eq.~(\ref{eqn:model}).  a) No Ag added to the \Agsqrt b) With 16 $10^{-3}$ ML of Ag adsorbed after the heating
cycles. \label{fig:gas}}
\end{figure}

For this purpose, we took a freshly prepared sample with 1.2 ML of Ag and carried out two 
heating cycles to 600\gradC. Then we performed EELS-LEED measurements at constant $k_\parallel$ with 
respect to the (00) spot, and restarted the measurement typically every hour, 
working at the base pressure in the system 
of $1\times 10^{-10}$ mbar. The shift of the plasmon loss peak as a function of time is shown in 
Fig.~\ref{fig:gas}a). In order to be able to plot several different $k_\parallel$
into the same graph, and to combine different runs, we normalized the data to the saturation value for each 
$k_\parallel$. Interestingly, the saturation value of the plasmonic loss obtained by residual gas 
adsorption,$E_0(k_\parallel)$, turned out to be very close to the value obtained after simple adsorption of 
1.0 ML of Ag  without heating the surface afterwards \cite{Krieg13}. 
Please note that time zero is taken when measurements were started, which does not coincide 
with the effective start of residual gas adsorption, since several control measurements were carried out 
before the actual measurement was started. The typical time constant, as deduced from the fit described 
below, was found to be about 5 hours, 
in qualitative agreement with expectations of residual gas adsorption at this pressure. 
It should be noted here that these findings are not due to any mobility of Ag atoms on the surface, as  
concluded from the controlled adsorption of Ag described below. 

In fact, the change of plasmon loss energy as seen in  Fig.~\ref{fig:gas}a) 
can be almost quantitatively modeled (see red lines in this figure) by assuming 
that an incoming atom is only adsorbed if it hits 
a free adsorption site directly, and is reflected otherwise (Langmuir adsorption). 
This is reasonable, since adsorption of 
the typical constituent of the residual gas (CO, H$_2$O, H$_2$) will not be adsorbed on the Ag-covered
(111) terraces, apart from defects. They may, however, be adsorbed at step sites. In any case, 
the time available for diffusion will be very short, so that direct adsorption is most likely. 
This model leads to a linear decay of the adsorption probability as a function of active site 
concentration, and an exponential saturation of the additional coverage with time.  
In order to come to a quantitative description of  the observed adsorption 
behavior, we have to make two further assumptions: a) The additionally adsorbed atoms make 
a constant charge transfer per particle into the surface state, $S_1$, and b) a square root like 
dependence between the plasmon energy and the electron density exists. while the first assumption can only 
be rationalized by the still low concentration of charge carriers (for an estimate, see below), 
the second is substantiated by further experiments with controlled amounts of Ag, described in the next 
section. We then obtain the 
following formula, which describes the time dependence of the plasmonic electron loss energy at constant 
$k_\parallel$, normalized to the respective saturation value:
\begin{equation}\label{eqn:model}
 \frac{E(\vec{k},t)}{E(\vec{k},\infty)} =: E(t) = \sqrt{\left(\Theta_0-1\right) \exp(-s\rho\cdot t) +1 }
\end{equation}

The normalization by the saturation energy at each $k_\parallel$ was done in order to remove 
the dependence on momentum $\vec{k_\parallel}$ and to be able to plot the data into a common graph, if the 
assumptions made above are fulfilled.  
$\rho$ represents the particle flux per unit area hitting the surface per time, and $s$ is the 
probability that a particle will stick on a previously empty site. 
$\Theta_0$ represents the fraction of already occupied sites at $t=0$.

As demonstrated by the red  curves in Figs.~\ref{fig:gas}a) and b), the data fall well on the red lines 
obtained from fits with eq.~\ref{eqn:model}, where different time constants for the changes of 
$E_{loss}(k_\parallel)$ were obtained, depending on the amount of Ag added to the freshly 
prepared \Agsqrt surface {\em after} the heating cycles. Already an amount of 0.016 monolayers 
(ML) of Ag increases the time constant of the plasmonic shift induced by 
residual gas by a factor of 10. Therefore, the influence of residual gas on the results 
of Ag doping had to be subtracted only for the smallest Ag concentrations. 

The reasonably good fit to the data seems to justify the assumptions made in the model leading 
to eq.~\ref{eqn:model}. It contains several aspects: Apart from an indirect proof 
of the $\sqrt{n_e}$ dependence of the plasmonic energy also in 1D, the model 
assumes, as mentioned, that the ``dopants'' are an immobile 
species adsorbed on defects and/or steps of the surface. Additional Ag atoms seem to compete for 
the same defect sites with  
residual gas particles and are able to effectively block them against residual gas. 
Taking the analogy to the results found on flat Si(111) 
\cite{Crain05}, Ag atoms are expected to act as electron donors. Since the excess charge density 
at saturation is comparable to that found in our previous work \cite{Krieg13}, 
namely $1.9\times 10^7$cm$^{-1}$ per wire, 
we can easily estimate the density of particles necessary to produce this effect, 
assuming that one elementary charge per particle is transferred:
The Ag-covered (111) terraces are 11 Si units wide. Along the wires we have an 
Si density of $2.6\times 10^7$cm$^{-1}$ per atomic row. The Ag density in the 
\Agsqrt phase is the same as for Si. Assuming that the line density of charges 
determined above is evenly distributed among the Ag atoms in the $\sqrt{3}$ phase, 
this corresponds to about 0.066 elementary charges per Ag atom on the (111) terrace. 
It is quite unlikely that the density of defects fixing the charge transferring particles 
within the  \Agsqrt wires is so high as to provide this charge density. On the other hand, 
a single atomic chain with the Si density at a step edges adjacent to (111) terraces may provide 
a density of sites that is sufficient for adsorbing the amount of particles necessary for doping.

The strong change in time constant by pre-doping with Ag indicates that Ag  passivates 
not only the occupied site but also its  nearby surrounding by 
repulsive interactions with residual gas particles. In fact, Ag atoms turned out to be highly mobile at 
room temperature on the \Agsqrt/Si(111) terraces \cite{Sato1999}. Therefore it is quite likely 
that these atoms are only trapped at the step edges, where they become immobilized. 
Attempts to identify the adsorbed species directly by characteristic vibrational
losses turned out not to be successful, presumably for reasons of low concentration. 

\subsection{Doping via adsorption of silver adatoms}\label{sec:Ag}
\begin{figure}[tb]
 \centering
 \subfigure[]{\includegraphics[width = 0.45\textwidth]{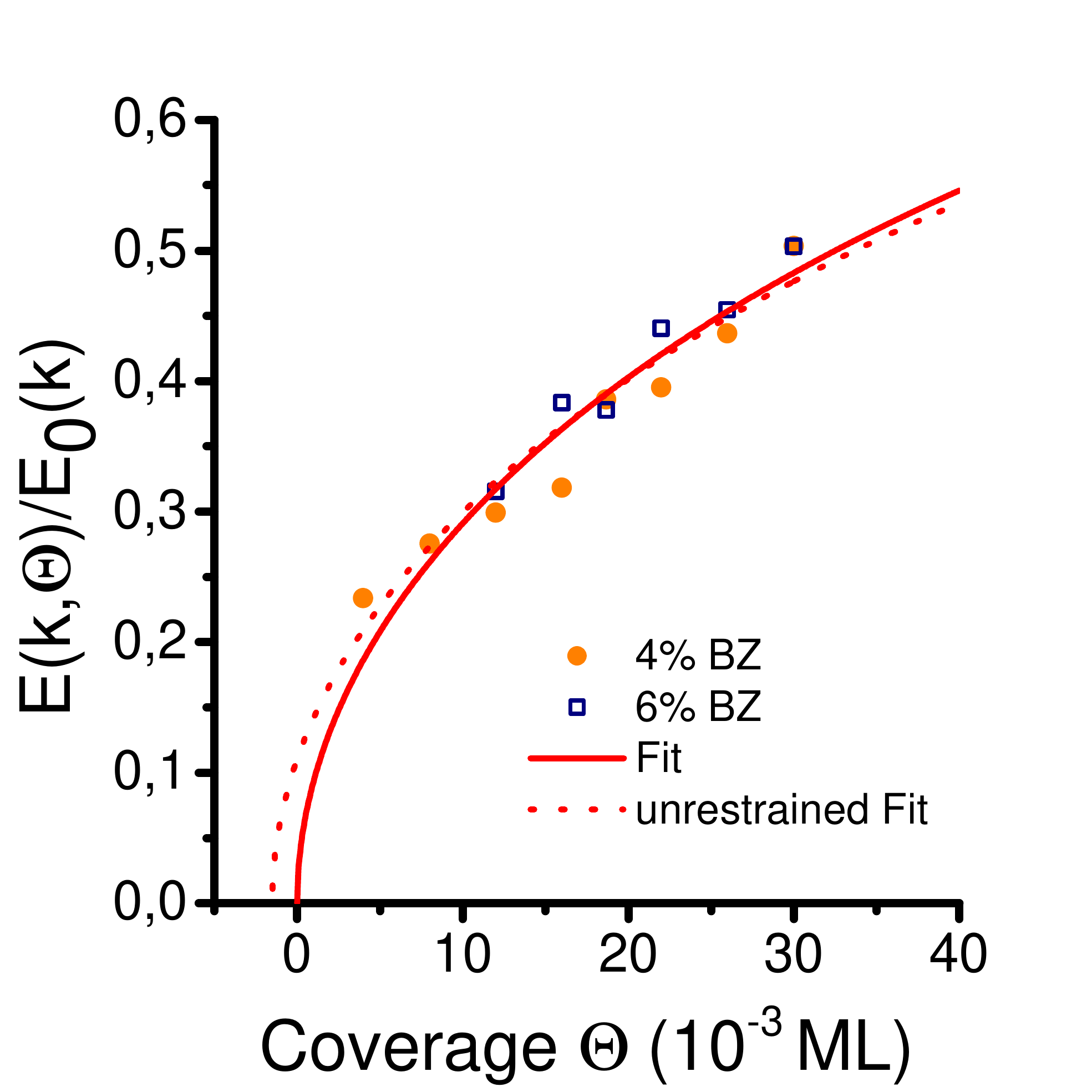}}\qquad
 \subfigure[]{\includegraphics[width = 0.45\textwidth]{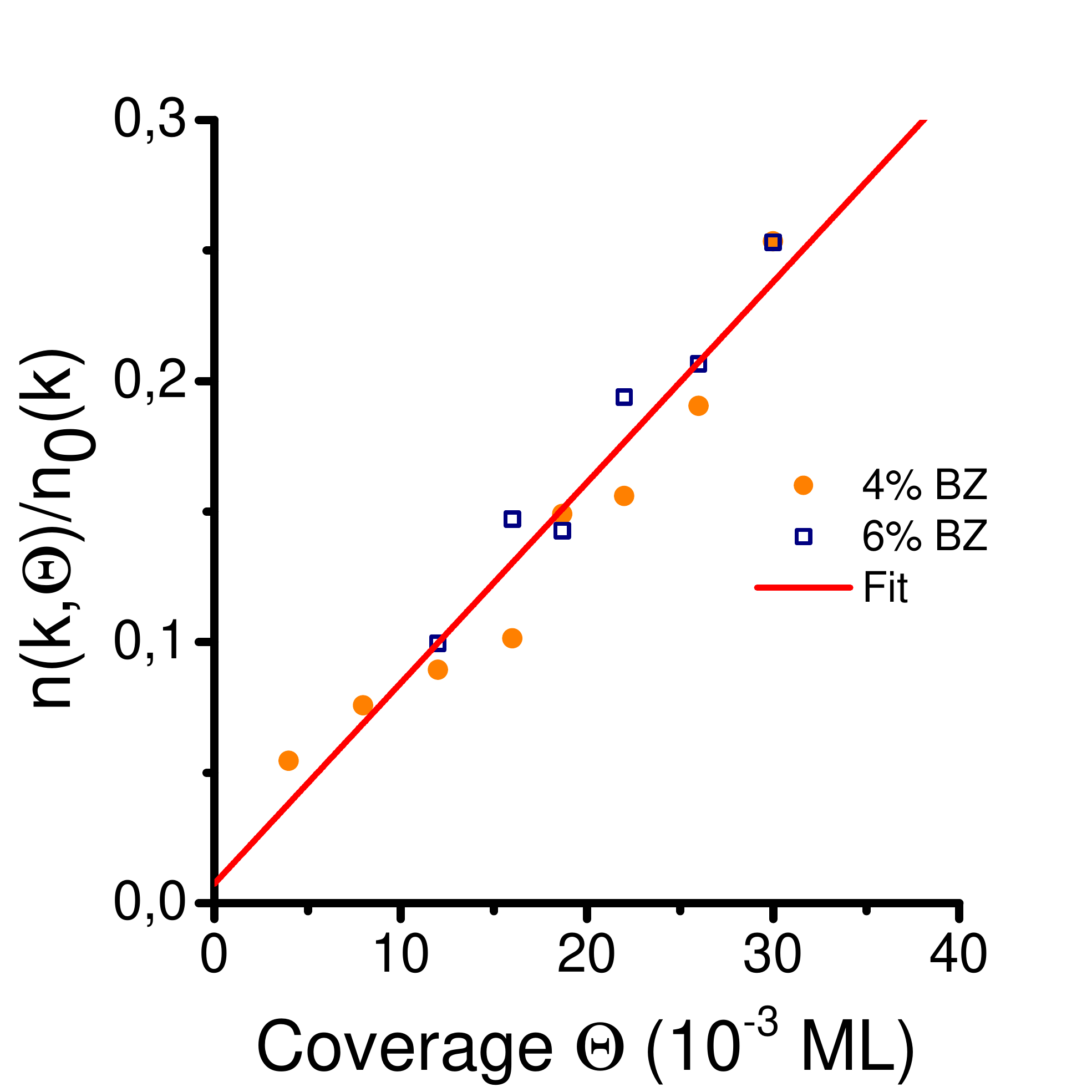}}
 \caption{Normalized plasmon losses as a function of Ag excess coverage. The orange dots and 
 the square were measured at 4\% BZ and 6\% BZ respectively. a) Data as obtained by silver doping corrected 
 for contributions from residual gas atoms. Red data: $\sqrt{\Theta}$-dependence without (dotted) and with restriction to start at $\Theta = 0$ on both axes.  b) Normalized 
 electron density. Red curve: Linear fit without restriction. Normalization was carried out with the value of ref.~\cite{Krieg13}. \label{fig:Doping}}
\end{figure}

While these experiments demonstrate the effect of charge transfer by adsorbed species qualitatively, we 
now go one step further by directly using the adsorption of Ag atoms on a \Agsqrt-ordered surface, 
prepared as described above, that shows no plasmonic losses. Here the amount adsorbed is well known due 
to our calibration. After preparation of 1.2 ML of Ag, we again carried out two heating cycles
and then evaporated various small amounts of silver with the sample held at RT, followed by 
EELS-LEED measurements at several scattering angles with respect to the (00) beam, i.e. at various $k_\parallel$.
Effects of residual gas adsorption have been carefully subtracted from the 
measured loss energies, but their influence could be kept small in all cases. 
In Fig.~\ref{fig:Doping}a) we show results of measured loss energies as a function of Ag exposure to the 
sample in units of $10^{-3}$~ML. In the small excess coverage range, 
$\Delta\Theta$, considered here (up to 3\% of a monolayer) the assumption of a constant sticking 
coefficient, which must be close to 1, seems to 
be well justified. Thus coverage and total charge carrier concentration are linearly related.  
Similar to the data treatment in 
Fig.~\ref{fig:gas}, we normalized the loss energies by the value obtained after preparation of 1.0 ML \cite{Krieg13}
 without further treatment for each $k_\parallel$ in order to combine data at different $k_\parallel$. 
Data at 4 and 6\% BZ are shown. As seen in Fig.~\ref{fig:Doping}a), 
the plasmonic energy losses are fully compatible 
with a $\sqrt{\Delta\Theta}$ behavior that goes through zero at zero excess coverage,
$\sqrt{\Delta\Theta}$, i.e. without offset. 

This means that
a) the \Agsqrt wires as prepared by our heating cycles are indeed semi-metallic.\\ 
b) The additional Ag atoms adsorbed afterwards are responsible for an effective charge transfer 
into the surface band, which gets partially filled due to the charge provided by the excess Ag atoms.\\ 
c) As seen in Fig.~\ref{fig:Doping}b), which plots the normalized square of a), a linear slope as a function of $\Delta\Theta$ 
is obtained. This is expected for a constant charge transfer per excess Ag atom and for a $\sqrt{n_e}$ dependence of the plasmonic loss energy. 
Therefore, we conclude that this functional dependence is also valid for 1D plasmons.

There is, of course, still the open question whether all adsorbed excess Ag atoms will contribute to 
the charge density in the surface band. Furthermore, it is unclear where these excess atoms are located. 
In order to answer these questions we carried out EELS measurements
at different temperatures between 80 and 400 K after preparation at RT. The plasmon losses generated 
by the excess Ag atoms turned out to be completely temperature independent. 
This finding rules out any phase equilibrium between a disordered lattice gas on top of the \Agsqrt wires
as suggested for the Ag/Si(111) system \cite{Liu09}.
It means that these excess atoms must be trapped and immobilized, i.e they are also chemisorbed quite
strongly. The adsorption must be  less strong than for the atoms in \Agsqrt wires, even at their edges, 
since we were able to preferentially desorb the excess atoms. Therefore, they cannot simply be edge atoms of the 
\Agsqrt wires. Most likely, they occupy sites at the steps adjacent to the \Agsqrt wires. Since 
bond formation there is different from the flat (111) terraces, dipole formation associated with charge
transfer is possible. This conclusion is consistent with the finding that 
the Ag atoms on top of the \Agsqrt wires are highly mobile at room temperature \cite{Hasegawa00}, 
so that they will be able to diffuse quickly into the stepped areas where they are trapped. The temperature 
independence of our findings over a very wide range of temperature indicate that thermal activation  
for diffusion plays no role. Therefore, we can safely assume that essentially all adsorbed excess Ag atoms 
contribute to the charge transfer leading to 1D plasmon formation. 

The induced electron density responsible for the plasmonic excitation can now be easily calculated 
via the change of the 1D plasmon energy for constant momentum 
\begin{equation}
 \frac{ E(\vec{k},\Theta)}{ E_0(\vec{k})} = \frac{\sqrt{n_e(\Theta)}}{\sqrt{n_{e,0}}}
\end{equation}
In Fig.~\ref{fig:Doping}b) we plot again the normalized values, 
where $n_{e,0}$ was again taken from the data of the 1.2ML Ag layer ($n_{e,0}=1.9\times 10^7$cm$^{-1}$).
As seen in Fig.~\ref{fig:Doping}b), we get an electron concentration of about 23\% of  $n_{e,0}$ by adsorbing
30 $10^{-3}$ ML of Ag. 

From the discussion above we concluded that the excess Ag atoms are concentrated on the steps of 
the (557) surface. There are two steps without direct contact to the \Agsqrt stripes. If these 
are the adsorption sites, we find that the homogeneous adsorption of 30 $10^{-3}$ ML of Ag on the whole 
surface, concentrated into two atomic chains, results in an average atomic density of 
$1.34\times 10^7$cm$^{-1}$ in the two chains. The 
electron density after excess doping by 30 $10^{-3}$ ML of Ag obtained from Fig.~\ref{fig:Doping}b) 
is $0.44\times 10^7$cm$^{-1}$. Thus we estimate the electron transfer rate to be around 
1/3 of an elementary charge per Ag atom. It is obvious that this number 
can only be taken as a preliminary estimate
that relies on assumptions still to be tested in more detail. Such investigations are in progress. 
\section{Summary and conclusions}
In this paper we gave evidence for the fact that, similar to the Ag/Si(111) system,  
also \Agsqrt wires on Si(557) are semi-metallic. The Ag-induced surface state can, however, 
be made conducting by appropriate charge transfer, which can be done by components of the residual gas, 
most likely water, and - more quantitatively - by adsorption of excess atoms of Ag on the \Agsqrt phase. 
A doped phase, and thus a metallic array of wires,  turns out to be always generated by the 
standard procedure of preparation. We thus proved a doping mechanism to be effective for a 
quasi-1D system. All excess particles leading to doping could be removed by thermal treatment of the 
sample.  

Our proposed mechanism for charge transfer is not the standard mechanism of intrinsic doping, since it 
is effective also with the same species as that which forms the wires. It relies on the variability of  
chemical bond formation in locally different environments with corresponding charge transfer into the 
bands formed by the majority species. Secondly, the doping generated by this mechanism is extrinsic on 
the atomic scale in the sense that the doping material is not incorporated into the Ag wires.  

Quantitative dosage of excess Ag atoms showed that the concentration of excess charges depends 
linearly on excess Ag concentration, if a $\sqrt{n_e}$ dependence of plasmonic loss energies is assumed,   
as predicted by theory \cite{Li90}. This theory, however, is based on the nearly-free electron gas model that is not expected to be valid in 1D, but this square root dependence on $n_e$ seems to survive. 

Estimates show that only a fraction 
of an elementary charge is transferred, as expected for a mostly covalent bond formed between 
Ag and Si. If other species like water or OH allow a higher charge transfer, the amounts 
necessary for the observed charge concentrations (only a few percent of a monolayer) may not 
easily be detectable by EELS for intensity reasons. 

\ack Financial support by the Nieders\"achsisches Ministerium f\"ur Wissenschaft
und Kultur through the graduate school ``Contacts in Nanosystems'' and by the Deutsche
Forschungsgemeinschaft is gratefully acknowledged.

\section*{References} 
\bibliographystyle{iopart-num}
\bibliography{low-d-plasmons.bib}

\end{document}